\documentclass[english]{article}
\usepackage[T1]{fontenc}
\usepackage[utf8]{inputenc}
\usepackage{mathrsfs}
\usepackage{amsmath}
\usepackage{amssymb}
\usepackage{graphicx}
\usepackage{babel}
\begin{document}
\title{Hidden Killing Fields, Geometric Symmetries and Black Hole Mergers}
\author{Albert Huber\thanks{hubera@technikum-wien.at}}
\date{{\footnotesize{}UAS Technikum Wien - Department of Applied Mathematics
and Physics, H\"ochst\"adtplatz 6, 1200 Vienna, Austria}}
\maketitle
\begin{abstract}
In the present work, using the recently introduced framework of local
geometric deformations, special types of vector fields - so-called
hidden Killing vector fields - are constructed, which solve the Killing
equation not globally, but only locally, i.e. in local subregions
of spacetime. Taking advantage of the fact that the vector fields
coincide locally with Killing fields and therefore allow the consideration
of integral laws that convert into exact physical conservation laws
on local scales, balance laws in dynamical systems without global
Killing symmetries are derived that mimic as closely as possible the
conservation laws for energy and angular momentum of highly symmetric
models. The utility of said balance laws is demonstrated by a concrete
geometric example, namely a toy model for the binary merger of two
extremal Reissner-Nordstr\"om black holes.
\end{abstract}
\textit{\footnotesize{}Key words: hidden Killing vectors, phantom
symmetries, conservation laws }{\footnotesize\par}

\section*{Introduction}

Due to a lack of continuous symmetry, the Killing equation cannot
be solved in generic geometric settings. Consequently, there are no
Killing vector fields in generic spacetimes. 

As a result, it turns out that said spacetimes are not endowed with
the geometric properties required for the formulation of physical
conservation laws for quantities such as energy, momentum and angular
momentum. 

However, this does not mean that without Killing vectors (KV) it would
be impossible to define globally conserved currents in generic spacetimes.
On the contrary, even in geometric models without Killing symmetries,
there is always an infinite number of such currents known as Komar
currents \cite{komar1959covariant}, which can be constructed from
any given vector field regardless of the concrete geometric structure
of spacetime. Furthermore, in special classes of spacetimes that lack
not all, but only specific Killing symmetries, there are special types
of vector fields, such as e.g. Kodama vector fields \cite{Kodama:1979vn}
in the case of spherically symmetric spacetimes, which lead to exact
conserved quantities and associated exact integral conservation laws.

In the literature, vector fields with the property that they permit
the definition of well-defined conservation laws in generic spacetimes
(same as the Kodama vector fields mentioned above) are often assigned
to the so-called class of generalized Killing vector fields (GKV).
The main property of the representatives of this class - of which
the most well-known are probably semi-Killing and almost Killing vector
fields as well as affine and curvature collineations \cite{beetle2014perturbative,bona2005almost,carot1994matter,Cook:2007wr,feng2018some,feng2019almost,garfinkle1999symmetry,katzin1969curvature,komar1959covariant,komar1962asymptotic,matzner1968almost,radhakrishna1984pure,ruiz2014almost,taubes1978solution}
- is that the vector fields in question are, in contrast to traditional
KV, not solutions of the Killing equation. Instead, said vector fields
represent solutions of less restrictive types of equations, which
can be solved under more general geometric circumstances, but which
still allow the construction of quantities with similar geometric
properties as those available in spacetime with isometries.

Common to all these types of GKV (including Kodama vectors) and their
associated conserved currents is that they are defined globally, i.e.
with respect to the global symmetry properties of spacetime. Therefore,
they prove to be parallel to - or exactly identical with - KV only
if the geometry of spacetime has global Killing symmetries.

As a consequence, however, the question arises whether there are generalized
KV in generic spacetimes that do not coincide globally but only locally
(i.e. in local subregions of spacetime) with solutions of the Killing
equation. This question is related to a concrete physical idea, namely
to the idea that there should exist classes of spacetimes whose geometries
change with time in such a way that their symmetry properties change
as well. To understand this, one may recall that it is eventually
to be expected that stationary physical systems (which are subject
to external disturbances) can become unstable over time or that certain
physical processes, such as gravitational collapse, may cause an initially
spherically symmetric geometry to lose some of its symmetry properties
and change e.g. to an axially symmetric geometry for a certain time.
In such a case, the local geometric structure of spacetime would allow
for the existence of corresponding KV, whilst its global structure
would not allow the same.

To account for this particular aspect of the theory and to show that
there are indeed classes of spacetimes that allow for the existence
of special types of GKV that coincide with exact KV on local scales,
to be referred to as \textit{hidden Killing vectors} (HKV) from now
on, the recently introduced framework of local geometric deformations
\cite{Huber:2019cze} is used in the present work. 

This geometric framework is well suited to prove the existence of
HKV for several reasons, the most important of which is certainly
that it allows the introduction of the concept of a so-called local
spacetime (i.e. a spacetime whose local geometric structure differs
from that of an associated ambient spacetime). By introducing this
particular type of spacetime, the geometric framework mentioned provides
a basis for the analysis of the local geometric structure of a given
spacetime with respect to its global structure, making it possible
(at least in principle) to model complicated geometric transitions
of local spacetimes whose symmetry properties change with time. 

The main idea to model such geometric transitions, as explained in
the first section of this work, is to deform background spacetimes
with certain Killing symmetries in such a way that they lose their
symmetry properties, yet using deformation fields with compact supports
that vanish in local subregions of spacetime. In this manner, it is
ensured by construction that spacetime exhibits HKV. But since spacetime
still does not allow for actual Killing symmetries, but only geometric
symmetries that correspond to Killing symmetries on local scales,
the term \textit{phantom symmetries} is used to indicate that one
is not dealing with actual symmetries, but rather with geometric (apparent)
symmetries that are confined to a Lorentzian submanifold of spacetime.

As shown in the rear part of the first section of this work, the existence
of spacetimes that allow HKV and associated phantom symmetries has
an interesting consequence, namely that certain integral laws of the
theory reduce to exact conservation laws, resulting in exact expressions
for conserved charges for energy and angular momentum on local scales.
By means of this insight, physical balance laws for energy and angular
momentum are derived in this work, which describe how exactly these
locally conserved quantities change over time in the course of the
development of a superordinate generic geometric model. The derived
balance laws have the interesting property that they are very closely
related to the conservation laws of known symmetric models and coincide
with these laws except for correction terms resulting from the geometric
deformation of the symmetric background geometry, thereby allowing
the generalization of the mentioned laws to more general geometric
settings.

To demonstrate the utility of said laws, a special geometric model
is treated in section two of this work, namely a toy model for a black
hole merger of two extremal Reissner-Nordstr\"om black holes. To provide
an exact characterization of the geometric structure of a merger of
this type, a non-stationary axisymmetric geometric model is considered
which coincides by construction locally with the (two-body) Majumdar-Papapetrou
solution at early times and the Reissner-Nordstr\"om solution at late
times. As is demonstrated, the geometric structure of the merger field
is consistent with Hawking's area theorem and, moreover, allows the
definition of Komar integrals at infinity, which may be used as a
starting point for the formulation of the laws of black hole mechanics
\cite{Bardeen:1973gs}. A key aspect, in this respect, turns out to
be that the (timelike) HKV of the geometry coincides with the (timelike)
KV of Majumdar-Papapetrou spacetime at early times and that of Reissner-Nordstr\"om
spacetime at late times. After demonstrating that the merger geometry
admits a HKV with these properties, a local Killing charge is defined
and an associated energy balance law is derived, which extends the
'standard' energy conservation laws of known symmetric black hole
models. A discussion of all the results obtained forms the conclusion
of this treatise.

\section{Local Spacetimes, Hidden Killing Fields and Phantom Symmetries}

The main objective of this section is to introduce the concept of
a\textit{ }HKV and the closely related concept of a phantom symmetry
and to show that these concepts can be used to derive balance laws
for energy and/or angular momentum in generic non-stationary spacetimes.
For the purpose of introducing these concepts, a pair of spacetimes
$(M,g)$ and $(\tilde{M},\tilde{g})$ with Lorentzian manifolds $M\subseteq\tilde{M}$
shall be considered, whose metrics are subject to the conditions 
\begin{equation}
\tilde{g}_{ab}\vert_{M}\equiv g_{ab},\;\tilde{g}^{ab}\vert_{M}\equiv g^{ab}.
\end{equation}
A specific way to meet these conditions, as recently pointed out in
\cite{Huber:2019cze}, is the use of the geometric framework of local
metric deformations. This framework deals with metric deformation
relations of the form

\begin{equation}
\tilde{g}_{ab}=g_{ab}+e_{ab}
\end{equation}
and
\begin{equation}
\tilde{g}^{ab}=g^{ab}+f^{ab},
\end{equation}
where the deformation tensor fields $e_{ab}$ and $f^{ab}$ are assumed
to be smooth tensor fields of compact support. In particular, $supp\,e_{ab},\,supp\,f^{ab}\equiv\tilde{M}\text{\ensuremath{\backslash}}M$
shall apply in this context, where a consistent choice is e.g. $e_{ab}=\chi\mathfrak{e}{}_{ab}$
with $\chi(x)$ being either an indicator function such as e.g. the
Heaviside step function or a smooth non-analytic transition function
of the form 
\begin{equation}
\chi(x,x_{0})=\frac{\psi(\frac{x}{x_{0}})}{\psi(\frac{x}{x_{0}})-\psi(1-\frac{x}{x_{0}})},
\end{equation}
which is constructed from a cut-off function 
\begin{equation}
\psi(x):=\begin{cases}
\overset{e^{-\frac{1}{x}}}{\underset{0}{}} & \overset{x>0}{\underset{x\leq0}{}}\end{cases}
\end{equation}
meeting the the conditions $0\leq\psi\leq1$ and $\psi(x)>0$ if and
only if $x>0$. Note that the point $x_{0}$ is introduced merely
to ensure that the exponent in $(5)$ is dimensionless. Of course,
a splitting like this can neither be unique nor coordinate-independent.
In fact, there are many different ways to realize splittings of the
form $(2)$ and $(3)$. However, assuming for simplicity that $g_{ab}$
and $\tilde{g}_{ab}$ are given in the same coordinates, the splitting
can be unambiguously performed. 

For consistency reasons, the metric $\tilde{g}_{ab}$ and and its
inverse $\tilde{g}^{ab}$ must fulfill the relations

\begin{equation}
\tilde{g}_{ab}\tilde{g}^{bc}=\delta_{a}^{\:c},
\end{equation}
which can be re-written in the form
\begin{equation}
e_{a}^{\:b}+f_{a}^{\:b}+e_{a}^{\:c}f_{c}^{\:b}=0,
\end{equation}
where $e_{a}^{\:b}=g^{bc}e_{ac}$ and $f_{a}^{\:b}=g_{ac}f^{bc}$.

Due to the fact that the deformation fields are chosen to be of compact
support, it turns out - in the event that consistency conditions $(7)$
are met - that the local expressions
\begin{equation}
g_{ab}=\tilde{g}_{ab}-e_{ab},g^{ab}=\tilde{g}^{ab}-f^{ab}
\end{equation}
are well-defined tensor fields on $\tilde{M}$, which coincide locally
(i.e. in $M$) with the metric $\tilde{g}_{ab}$ and the inverse metric
$\tilde{g}^{ab}$ of spacetime. For that reason mainly, the pair $(M,g)$
should not be viewed as a 'proper' spacetime, but rather as a \textit{local
spacetime}, or, more precisely, as a spacetime whose associated Lorentzian
manifold $M$ is contained (and therefore localized) in another Lorentzian
manifold $\tilde{M}$ of the \textit{ambient spacetime} $(\tilde{M},\tilde{g})$.
Ultimately, the use of the term \textit{local} can be justified in
this context by the fact that $(M,g)$ and $(\tilde{M},\tilde{g})$
are spacetimes, whereas $(\tilde{M},g)$, in contrast, is not. 

Using decomposition relations $(2)$ and $(3)$, one can calculate
the difference tensor 
\begin{align}
C_{\,bc}^{a} & =\frac{1}{2}\tilde{g}^{ad}(\nabla_{b}\tilde{g}_{dc}+\nabla_{c}\tilde{g}_{bd}-\nabla_{d}\tilde{g}_{bc})=\\
 & =\frac{1}{2}(g^{ad}+f^{ad})(\nabla_{b}e_{dc}+\nabla_{c}e_{bd}-\nabla_{d}e_{bc})\nonumber 
\end{align}
which can be obtained by taking the difference of the Levi-Civita
connections of $(M,g)$ and $(\tilde{M},\tilde{g})$, meaning that
$C_{\,bc}^{a}=\tilde{\Gamma}{}_{\,bc}^{a}-\Gamma_{\,bc}^{a}$. This
difference tensor can be used to decompose the Riemann tensor of $(\tilde{M},\tilde{g})$
in the following way

\begin{equation}
\tilde{R}_{\,bcd}^{a}=R_{\,bcd}^{a}+E_{\,bcd}^{a},
\end{equation}
where $E_{\,bcd}^{a}=2\nabla_{[c}C_{\,d]b}^{a}+2C_{e[c}^{a}C_{d]b}^{e}$
holds by definition. By contracting indices, the decomposition 

\begin{equation}
\tilde{R}_{bd}=R_{\,bd}+E_{bd}
\end{equation}
Ricci tensor is obtained, where, of course, $E_{bd}=\delta_{a}^{\:c}E_{\,bcd}^{a}=2\nabla_{[a}C_{\,d]b}^{a}+2C_{e[a}^{a}C_{d]b}^{e}$
must apply. By repeating this procedure, the corresponding relation 

\begin{equation}
\tilde{R}=R+g^{bd}E_{bd}+f^{bd}R_{bd}+f^{bd}E_{bd}
\end{equation}
for the Ricci scalar is obtained, which, however, allows one to decompose
Einstein's field equations 

\begin{equation}
\tilde{G}_{ab}=8\pi\tilde{T}_{ab}
\end{equation}
in the form

\begin{equation}
G_{ab}+\rho_{ab}=8\pi\tilde{T}_{ab}.
\end{equation}
Using now the fact that the local field equations 
\begin{equation}
G_{ab}=8\pi T_{ab}
\end{equation}
are met in $(M,g),$ the \textit{deformed Einstein equations}

\begin{equation}
\rho_{ab}=8\pi\tau_{ab}
\end{equation}
can be set up, provided that $\tau_{ab}=\tilde{T}_{ab}-T_{ab}$.

Given this geometric setting, a HKV is given when a continuous symmetry
of $g_{ab}$ is generated by a Killing vector $\zeta^{a}$ (throughout
$M$), which is not a (global) continuous symmetry to $\tilde{g}_{ab}$,
so that

\begin{equation}
L_{\zeta}g_{ab}=2\nabla_{(a}\zeta_{b)}=0,\;L_{\zeta}\tilde{g}_{ab}=2\tilde{\nabla}_{(a}\zeta_{b)}\neq0
\end{equation}
applies due to the fact that $L_{\zeta}e_{ab}\neq0$, where $L_{\zeta}$
is the Lie derivative along $\zeta^{a}$. More specifically, a spacetime
$(\tilde{M},\tilde{g})$ has a HKV with respect to $(M,g)$ if there
exist local tensor fields $e_{ab}$ and $f^{ab}$ such that conditions
$(1)$ and $(17)$ are met and therefore a vector field $\zeta^{a}$
which has the Killing property only with respect to the $(M,g)$,
but not with respect to $(\tilde{M},\tilde{g})$\footnote{Regarding this definition, it is important to keep in mind that a
HKV and an associated hidden symmetry for $g_{ab}$ is always given
relative to a specific choice for the ambient metric $\tilde{g}_{ab}$.
However, this choice of the ambient metric $\tilde{g}_{ab}$ is generally
not unique, since a multitude of possible ambient spacetimes $(\tilde{M},\tilde{g})$
exist for a local spacetime $(M,g)$.}. 

The present definition of a HKV is closely related to the definition
of a GKV\footnote{Note that various types of definitions of GKV have been given in the
literature over the years and that a specific one is considered at
this stage.} given in \cite{Harte:2008xt,Harte:2008xq}. Given in relation to
a timelike worldline $\mathcal{Z}$ with unit tangent $w^{a}$, a
GKV in the mentioned works is defined by the relations
\begin{equation}
L_{\zeta}g_{ab}\vert_{\mathcal{Z}}=\nabla_{c}L_{\zeta}g_{ab}\vert_{\mathcal{Z}}=0
\end{equation}
and proven to be constructible by using Jacobi vector fields, i.e.
solutions of the geodesic deviation equation
\begin{equation}
\ddot{\zeta}^{a}=\tilde{R}_{\;bcd}^{a}w^{b}w^{c}\zeta^{d}
\end{equation}
with the property that 
\begin{equation}
L_{w}\zeta^{a}=0,
\end{equation}
where the overdot is short hand notation for the total derivative
$(w\tilde{\nabla})=w^{a}\tilde{\nabla}_{a}$. An important feature
of these GKV is the following: If $(\tilde{M},\tilde{g})$ has Killing
symmetries, it turns out that every GKV of the above form is always
an ordinary KV (but, of course, not vice versa).

Although not really obvious at first glance, the definition of a HKV
given above is quite similar to that of a GKV, as the geometric setting
discussed requires that

\begin{equation}
L_{\zeta}\tilde{g}_{ab}\vert_{M}=\tilde{\nabla}_{c}L_{\zeta}\tilde{g}_{ab}\vert_{M}=0
\end{equation}
applies. The main difference between the two definitions is therefore
that both the Lie derivative of the metric and its covariant derivative
must be zero throughout $M$ and not exclusively along a single timelike
worldline $\mathcal{Z}$. However, since $\zeta^{a}$ must be a KV
by construction along any curve (or curve segment) in $M\subseteq\tilde{M}$,
it is, of course, also a GKV.

Anyway, due to the imposed conditions on the geometric structure of
$e_{ab}$ and $f^{ab},$ it becomes clear that there is a locally
conserved current $j^{a}=-G_{\;b}^{a}\zeta^{b}$ being subject to
the conservation law 
\begin{equation}
\nabla_{a}j^{a}=0,
\end{equation}
which, however, holds in this form only in $M$, i.e. the submanifold
of $\tilde{M}$ in which spacetime geometry becomes symmetric.

As a direct result, considering a submanifold $\mathcal{D}\subset M$
with boundary $\partial\mathcal{D}$, which is contained in a larger
submanifold $\tilde{\mathcal{D}}\subset\tilde{M}$ with boundary $\partial\mathcal{\tilde{D}}$
such that $\mathcal{D}\subseteq\tilde{\mathcal{D}}$, the Gaussian
theorem can be used when integrating relation $(22)$ to obtain

\begin{equation}
\underset{\mathcal{\mathcal{D}}}{\int}\nabla_{a}j^{a}dv=\underset{\partial\mathcal{\mathcal{D}}}{\int}j^{a}d\Sigma_{a},
\end{equation}
where $dv\equiv\sqrt{-g}d^{4}x$ and $d\Sigma_{a}=n_{a}d\sigma$ with
$d\sigma\equiv\sqrt{h}d^{3}x$ are the corresponding four-volume and
induced hypersurface elements, respectively. 

On the other hand, by defining the further current $\tilde{j}^{a}=-\tilde{G}_{\;b}^{a}\zeta^{b}$
and integrating
\begin{equation}
\tilde{\nabla}_{a}\tilde{j}^{a}=\tilde{G}_{\;b}^{a}\tilde{\nabla}_{a}\zeta^{b},
\end{equation}
one obtains the result

\begin{equation}
\underset{\mathcal{\tilde{D}}}{\int}\tilde{\nabla}_{a}\tilde{j}^{a}d\tilde{v}=\underset{\mathcal{\tilde{D}}}{\int}G_{\;b}^{a}\tilde{\nabla}_{a}\zeta^{b}d\tilde{v},
\end{equation}
which holds throughout $\tilde{\mathcal{D}}$ and reduces to the form
$(23)$ in $\mathcal{\mathcal{D}}.$ Consequently, however, the LHS
of this relation can be re-written in the form 
\begin{equation}
\underset{\mathcal{\tilde{D}}}{\int}\tilde{\nabla}_{a}\tilde{j}^{a}d\tilde{v}=\underset{\mathcal{\mathcal{D}}}{\int}\nabla_{a}j^{a}dv+\underset{\mathcal{\tilde{D}}\backslash\mathcal{D}}{\int}\tilde{\nabla}_{a}\tilde{j}^{a}d\tilde{v},
\end{equation}
and be converted to 
\begin{equation}
\underset{\partial\mathcal{\tilde{D}}}{\int}\tilde{j}^{a}d\tilde{\Sigma}_{a}=\underset{\partial\mathcal{\mathcal{D}}}{\int}j^{a}d\Sigma_{a}+\underset{\partial\{\mathcal{\tilde{D}}\backslash\mathcal{D}\}}{\int}\tilde{j}^{a}d\tilde{\Sigma}_{a}.
\end{equation}
after repeated application of the Gaussian theorem.

These results, although they appear not to be particularly special
or interesting at first sight, have nice applications under special
geometric circumstances, i.e. when the ambient spacetime $(\tilde{M},\tilde{g})$
is a sandwich spacetime with stationary and axisymmetric initial and
final geometries $(M,g)$ and $(M',g')$ and a dynamical transition
spacetime $(\mathcal{O},\tilde{g})$ connecting both of these local
spacetimes such that the manifold structure is $\tilde{M}=M\cup\mathcal{O}\cup M'$:
They allow, as shall be demonstrated in the following, the derivation
of balance relations that generalize the conservation laws for energy
and angular momentum holding with respect to the symmetric parts of
spacetime, that is, with respect to the initial and final geometric
configurations $(M,g)$ and $(M',g')$ of the generic spacetime $(\tilde{M},\tilde{g})$. 

To prove this assertion, it shall therefore now be assumed that the
Lorentzian manifold of $(\tilde{M},\tilde{g})$ is of the form $\tilde{M}=M\cup\mathcal{O}\cup M'$
and that the metric of spacetime $\tilde{g}_{ab}$ and its inverse
$\tilde{g}^{ab}$ locally meet the conditions

\begin{equation}
\tilde{g}_{ab}\vert_{M}\equiv g_{ab},\;\tilde{g}^{ab}\vert_{M}\equiv g^{ab}
\end{equation}
and

\begin{equation}
\tilde{g}_{ab}\vert_{M'}\equiv g'_{ab},\;\tilde{g}^{ab}\vert_{M'}\equiv g'^{ab},
\end{equation}
where $g_{ab}$ and $g'_{ab}$ are the metrics of two stationary and
axisymmetric local spacetimes $(M,g)$ and $(M',g')$. The listed
conditions can be met by using the same methods as above, namely by
considering the metric deformation relations

\begin{equation}
\tilde{g}_{ab}=g_{ab}+e_{ab}=g'_{ab}+e'_{ab}
\end{equation}
and
\begin{equation}
\tilde{g}^{ab}=g^{ab}+f^{ab}=g'^{ab}+f'^{ab},
\end{equation}
and by requiring that $supp\,e_{ab},\,supp\,f^{ab}\equiv\tilde{M}\text{\ensuremath{\backslash}}M$
and $supp\,e'_{ab},\,supp\,f'^{ab}\equiv\tilde{M}\text{\ensuremath{\backslash}}M'$
; geometric conditions that can be met by making the choices $e_{ab}=\chi\mathfrak{e}{}_{ab}$
and $e'_{ab}=\chi\mathfrak{e}'{}_{ab}$ for the corresponding deformation
fields in $(30)$, where $\chi=\chi(x)$ is a transition function
of the form $(4)$.

Based on the fact that these specific geometric conditions are met,
it is guaranteed that the generic spacetime $(\tilde{M},\tilde{g})$
allows for the existence of two HKV $\zeta^{a}$ and $\zeta'^{a}$
such that 

\begin{equation}
L_{\zeta}g_{ab}=2\nabla_{(a}\zeta_{b)}=0=2\nabla'_{(a}\zeta'_{b)}=L_{\zeta'}g'_{ab},\;L_{\zeta}\tilde{g}_{ab}=L_{\zeta'}\tilde{g}_{ab}\neq0.
\end{equation}
Here, as shall be demonstrated in the next section by a concrete geometric
example, it may very well occur that the different HKV $\zeta^{a}$
and $\zeta'^{a}$ coincide (not in all, but in certain cases of interest),
so that $\zeta^{a}$ turns out to be the only HKV of the geometry.
But this is a special and not the general case, which is why, to continue
the main line of argument, two different currents $\tilde{j}^{a}=-\tilde{G}_{\;b}^{a}\zeta^{b}$
and $\tilde{j}^{'a}=-\tilde{G}_{\;b}^{a}\zeta^{'b}$ shall be defined,
which are conserved currents in $M$ and in $M'$, so that
\begin{equation}
\nabla_{a}j^{a}=0
\end{equation}
applies in $M$ and

\begin{equation}
\nabla'_{a}j'^{a}=0
\end{equation}
in $M'$. Considering then - similar to the above - two submanifolds
$\mathcal{D}\subset M$ and $\mathcal{D}'\subset M'$ with boundaries
$\partial\mathcal{D}$ and $\partial\mathcal{D}'$, which are contained
in an associated ambient submanifold $\tilde{\mathcal{D}}\subset\tilde{M}$
with boundary $\partial\mathcal{\tilde{D}}$ such that $\mathcal{D},\mathcal{D}'\subseteq\tilde{\mathcal{D}}$,
one finds by repeating the formal steps described above

\begin{align}
\underset{\partial\mathcal{\tilde{D}}}{\int}\tilde{j}^{a}d\tilde{\Sigma}_{a} & =\underset{\partial\mathcal{\mathcal{D}}}{\int}j^{a}d\Sigma_{a}+\underset{\partial\{\mathcal{\tilde{D}}\backslash\mathcal{D}\}}{\int}\tilde{j}^{a}d\tilde{\Sigma}_{a}=\\
 & =\underset{\partial\mathcal{\mathcal{D}}'}{\int}j'^{a}d\Sigma'_{a}+\underset{\partial\{\mathcal{\tilde{D}}\backslash\mathcal{D}'\}}{\int}\tilde{j}^{a}d\tilde{\Sigma}_{a}.\nonumber 
\end{align}
\begin{figure}
\includegraphics[scale=0.3]{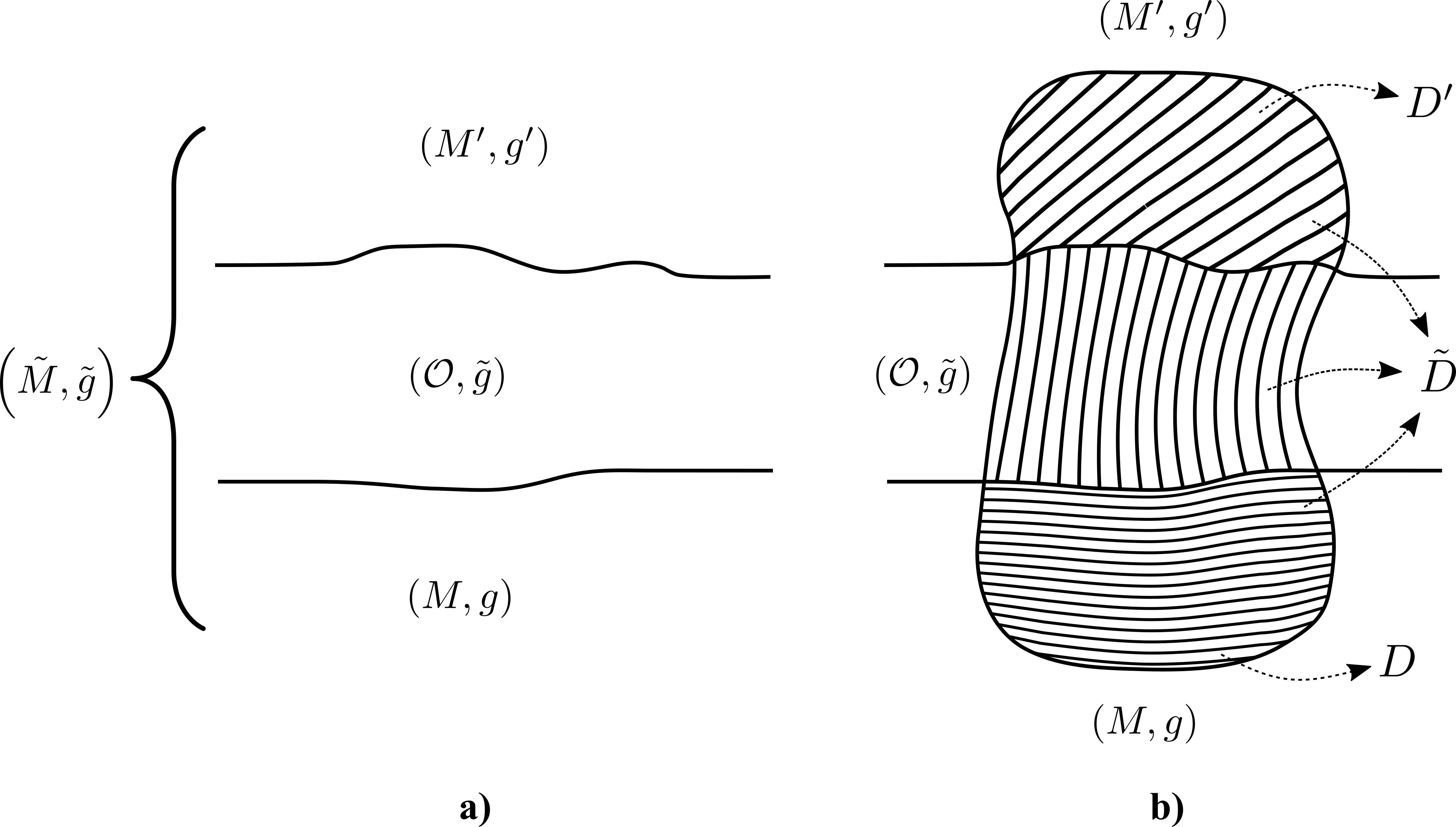}\caption{In graphic $\textbf{a)}$ to the left the geometric structure of the
ambient spacetime $(\tilde{M},\tilde{g})$ is depicted. To obtain
a sandwich spacetime $(\tilde{M},\tilde{g})$ of this kind, it is
assumed that the 'four-geometric' initial data $(M,g)$ and $(M'g')$
are stationary and axisymmetric local spacetimes in the sense of \cite{Huber:2019cze},
i.e. spacetimes with independent geometric structures and individual
symmetry properties, which do not necessarily have to coincide with
those of the ambient spacetime $(\tilde{M},\tilde{g})$. The graphic
$\textbf{b)}$ to the right shows how a submanifold $\tilde{\mathcal{D}}$
decomposes into three parts based on the decomposition $\tilde{M}=M\cup\mathcal{O}\cup M$,
where $\mathcal{D}$ lies in $M$ and $\mathcal{D}$' lies in $M'$.}
\end{figure}
This relation may be rewritten more compactly in the form
\begin{equation}
\tilde{Q}=Q+\mathcal{Q}=Q'+\mathcal{Q}',
\end{equation}
where $\tilde{Q}=\underset{\partial\mathcal{\tilde{D}}}{\int}\tilde{j}^{a}d\tilde{\Sigma}_{a}$,
$Q=\underset{\partial\mathcal{\mathcal{D}}}{\int}j^{a}d\Sigma_{a}$,
$Q'=\underset{\partial\mathcal{\mathcal{D}}'}{\int}j'^{a}d\Sigma'_{a}$
and $\mathcal{Q}=\underset{\partial\{\mathcal{\tilde{D}}\backslash\mathcal{D}\}}{\int}\tilde{j}^{a}d\tilde{\Sigma}_{a}$,
$\mathcal{Q}'=\underset{\partial\{\mathcal{\tilde{D}}\backslash\mathcal{D}'\}}{\int}\tilde{j}^{a}d\tilde{\Sigma}_{a}$.
Thus, provided that the definition $\mathscr{G}:=\mathcal{Q}-\mathcal{Q}'$
is now used in the present context, one obtains the balance law

\begin{equation}
Q'=Q+\mathscr{G}
\end{equation}
from these two decompositions of one and the same quantity $\tilde{Q}$. 

Consequently, considering now the case where spacetime admits only
one HKV for both local spacetimes $(M,g)$ and $(M',g')$, there is
an interesting conclusion that can be drawn from this derivation of
balance law $(37)$: Since $\zeta^{a}\equiv\zeta'^{a}$ may be chosen
both in $M$ and in $M'$ to be a linear combination of a timelike
and an axial Killing vector field such that $Q$ and $Q'$ are actual
Killing charges associated with the stationary, axisymmetric local
spacetimes $(M,g)$ and $(M',g')$ that are constant in time, it can
be concluded that also the derived quantity $\mathscr{G}$, which
measures how strongly both types of charges differ from each other,
must be constant in time. Formally expressed this means that - on
account of the fact that $L_{\zeta}Q=L_{\zeta}Q'=0$ applies in the
given case - the conclusion can be drawn that $L_{\zeta}\mathscr{G}=0$
must apply as well as a direct consequence.

As a result, however, despite being formulated in a generic geometric
setting, it is found that the balance law $(37)$ closely resembles
an actual integral conservation law for energy and/or angular momentum
charges of highly symmetric geometric models. This is not least because
$Q$ and $Q'$ are genuine conserved quantities in $M$ and $M'$,
respectively, which exist due to the fact that the local spacetimes
$(M,g)$ and $(M',g')$ have Killing symmetries, in stark contrast
to the ambient spacetime $(\tilde{M},\tilde{g})$, which has no such
symmetries at all. However, since the ambient spacetime $(\tilde{M},\tilde{g})$
has no exact symmetries whatsoever, relation $(37)$ cannot be an
exact physical conservation law, but rather has to be considered a
physical balance law, which relates conserved quantities existing
before and after the transition from one geometric configuration of
spacetime to another takes place.

In view of these results, it becomes clear that any spacetime $(\tilde{M},\tilde{g})$
with suitable four-geometric initial data, i.e. with stationary and
axisymmetric initial and final geometric configurations $(M,g)$ and
$(M',g')$, which exhibits one and the same HKV for both of these
local geometric configurations, allows for the definition of an apparent
conservation law of the form $(37).$ Since the occurrence of an exact
conservation law is always accompanied by the occurrence of a corresponding
exact symmetry, in the following, whenever a sandwich spacetime with
the above properties is given, it shall be said to have a \textit{phantom
symmetry} to indicate that spacetime in such a case has no actual,
but only apparent geometric symmetries, which arise as a result of
imposing special conditions on the geometry of spacetime at early
and late times.

Having said that, the next point to be stressed is that there are
different types of conserved currents $\tilde{j}^{a}$ that can be
considered in $(\tilde{M},\tilde{g})$. Some of these currents are
not only locally but globally conserved (in contrast to the types
of currents considered above). For the purposes of the present work,
however, only one type of conserved current will play a role, namely
the Komar current 
\begin{equation}
\tilde{j}^{a}=\tilde{\nabla}_{b}\tilde{\nabla}^{[b}\zeta^{a]},
\end{equation}
which can be constructed on any given spacetime $(\tilde{M},\tilde{g})$
for any given choice for the vector field $\zeta^{a}$. Given a generic
spacetime $(\tilde{M},\tilde{g})$ with a geometry that allows for
the existence of a phantom symmetry, that is, with stationary and
axisymmetric initial and final geometric configurations $(M,g)$ and
$(M',g')$ and a single HKV for both of these local configurations,
one may simply select $\zeta^{a}$ to construct such a Komar current
of the form $(38)$. Considering then a submanifold $\tilde{\mathcal{C}}\subset\tilde{M}$
with boundary $\partial\mathcal{\tilde{C}}$, which in turn is bounded
(typically at infinity) by a two-surface $\partial\tilde{\mathcal{C}}$,
the fact that $\tilde{\nabla}^{[b}\zeta^{a]}$ is antisymmetric can
be used to deduce from $(38)$ the conservation law
\begin{equation}
\underset{\tilde{\mathcal{C}}}{\int}\tilde{j}^{a}d\tilde{\Sigma}_{a}=2\underset{\partial\tilde{\mathcal{C}}}{\int}\tilde{\nabla}^{[a}\zeta^{b]}d\tilde{\Sigma}_{ab},
\end{equation}
where $d\tilde{\Sigma}_{ab}$ and $d\tilde{\Sigma}_{a}$ are the surface
and hypersurface elements of $\partial\tilde{\mathcal{C}}$ and $\tilde{\mathcal{C}}$,
respectively. Considering two further submanifolds $\mathcal{C}\subset M$
and $\mathcal{C}'\subset M'$ with boundaries $\partial\mathcal{C}$
and $\partial\mathcal{C}'$, which are contained in $\tilde{\mathcal{C}}$
such that $\mathcal{C},\mathcal{C}'\subseteq\tilde{\mathcal{C}}$,
the RHS of $(39)$ reads $\underset{\mathcal{C}}{\int}j^{a}d\Sigma_{a}=-\underset{\mathcal{C}}{\int}R_{\;b}^{a}\zeta^{b}d\Sigma_{a}$
in $M$ and $\underset{\mathcal{C'}}{\int}j^{'a}d\Sigma'_{a}=-\underset{\mathcal{C}}{\int}R_{\;b}^{'a}\zeta^{b}d\Sigma'_{a}$
in $M'$. Based on this input, a balance law of the form $(37)$ can
also be derived in the given case, whereas, of course, the resulting
charges are Komar charges in this particular case. In particular,
this allows the definition of the Komar expressions for energy and
angular momentum and the conclusion that those are conserved in the
local regions where spacetime becomes symmetric. Consequently, however,
the considered spacetime model is such that locally conservation of
energy and angular momentum holds, that is, specifically at early
and late times, if it has a hidden (phantom) symmetry.

Now that this all is clarified, the natural question arises what new
information can be obtained about the local metrics $g_{ab}$ and
$g'_{ab}$ and their associated ambient metric $\tilde{g}_{ab}$ now
that it is known that spacetime has a HKV and an associated hidden
symmetry? 

To answer this question, a brief detour shall be made and the theory
of extended irreversible thermodynamics \cite{israel1976nonstationary,israel1989covariant,israel1979transient,landau1987fluid}
shall be taken into account. In the mentioned theory, which typically
deals with the physical behavior of fluid mixtures near thermal equilibrium,
three different quantities play a central role, namely the so-called
thermal energy-momentum tensor $T_{\;b}^{a}=\epsilon w^{a}w_{b}+\varpi^{a}w_{b}+w^{a}\varpi_{b}+(p+\Pi)h{}_{\,b}^{a}+\Pi{}_{\,b}^{a}$,
the so-called particle number current $N_{A}^{a}=n_{A}w^{a}+\nu_{A}^{a}$,
defined with respect to a number $A$ of species of particles, and
the so-called covariant entropy current $S^{a}=sw^{a}+\eta^{a}$.
In this context, the scalar fields $\epsilon$, $p$ and $s$ represent
the energy density, pressure and total entropy density of the fluid
mixture, $n_{A}$ is the particle density of a number of particle
species $A$ and $\pi$ represents the viscous bulk pressure. The
vector fields $w^{a}$, $\varpi^{a},$ $\nu_{A}^{a}$ and $\eta^{a}$,
on the other hand, represent the four-velocity vector of the system,
an energy flux current and so-called particle diffusion and heat conduction
fluxes, while the tensor field $\pi{}_{\,b}^{a}$ represents the trace-free
anistropic viscous stress-tensor. As may be noted, there are different
ways to make an ansatz for the vector fields $\varpi^{a},$ $\nu_{A}^{a}$
and $\eta^{a}$, depending on whether one wants to specify them in
the so-called energy or particle reference frames. With respect to
one of these choices, the physical behavior of the fluid mixture can
be specified the following differential laws

\begin{equation}
\nabla_{a}T_{\;b}^{a}=0,\:\nabla_{a}N_{A}^{a}=0,\:\nabla_{a}S^{a}\geq0.
\end{equation}
The respective laws are the laws of local conservation of energy and
particle number and the so-called Clausius-Duhem inequality, which
is the differential form of the entropy law of thermodynamics. 

The extent to which all this matters for the question raised above
of what can be learned about the local and global metrics $g_{ab}$,
$g'_{ab}$ and $\tilde{g}_{ab}$ by means of hidden symmetries can
now be inferred from the following observation: To be in thermal equilibrium,
the matter distribution under consideration must satisfy special equilibrium
conditions, where the main equilibrium condition for relativistic
viscous fluids in extended irreversible thermodynamics (as well as
in classical theory) to be in thermal equilibrium is $\nabla_{a}S^{a}=0$.
For this condition to be fulfilled and for the corresponding system
to actually reach a local thermostatic equilibrium state, the heat
exchange between the fluids must cease and the sum of all thermal
potentials and chemical reaction rates must be zero. Furthermore,
the thermal energy-momentum tensor as well as the particle number
and entropy fluxes of the fluid mixture must match those of an ordinary
ideal fluid, which means that $T_{\;b}^{a}=(\epsilon+p)w^{a}w_{b}+p\delta{}_{\,b}^{a}$,
$N_{A}^{a}=n_{A}\cdot w^{a}$ and $S^{a}=s\cdot w^{a}$. For this
to hold, an equilibrium equation of state of the form $\epsilon+p=T(s+\underset{A}{\sum}\Theta{}_{A}n_{A})$
must be fulfilled, where each $\Theta_{A}$ represents a thermal potential
associated with a species $A$ of particles (characterized in terms
of the chemical potential $\mu_{A}\equiv\frac{\partial\epsilon}{\partial n_{A}}$).
Ultimately, however, and this is the crucial point, the motion of
the fluids should be rigid in Born' s sense, which means means that
the shear tensor constructed from the four-velocity $w^{a}$ must
vanish, i.e. $\sigma_{ab}=h_{a}^{\:c}h_{b}^{\:d}\nabla_{(c}w_{d)}=0$,
which in turn means that $\nabla_{(a}w_{b)}=0$ must be satisfied.
However, from this it follows that the matter field under consideration
must, at perfect thermal equilibrium, curve spacetime in such a way
that a stationary gravitational field is generated with a timelike
Killing vector field $\xi^{a}$ that is directly proportional to the
four-velocity $w^{a}$, i.e. $\xi^{a}=\beta w^{a}$ with $\beta\equiv c\cdot N\equiv c\cdot\sqrt{-\xi_{a}\xi^{a}}$
with $c=const.$ 

Conditions similar to the above must also be satisfied also in more
extended theories of irreversible thermodynamics \cite{hiscock1983stability,hiscock1985generic,jou1999extended,lindblom1983stability,maartens1996causal},
which avoid causality violations by requiring causal propagation of
dissipations. Eventually, even when considering types of matter fields
other than viscous fluids, it should be reasonable to expect that
the realization of thermodynamic equilibria (or rather quasi-equilibria)
requires the satisfaction of very similar, or at least comparable,
equilibrium conditions and thus the validity of the same type of stationarity
of the metric. 

Therefore, even though the above results of the theory of extended
irreversible thermodynamics may not be applicable to all geometric
models of general relativity, it can nevertheless be plausibly concluded
from them that in at least some cases of interest a given matter distribution,
in order to reach a thermal equilibrium state, must curve spacetime
in such a way that its geometry is (effectively) stationary and thus
exhibits time translation symmetry. However, this is only possible
if the field of said matter distribution hardly fluctuates, i.e.,
only in situations where the matter source is not subject to irreversible
dissipative processes, such as occur for example in the collision
of celestial bodies (or collections of them) or in the continuous
accretion of matter by a celestial body over long periods of time. 

Conversely, since it is an observational fact that compact massive
objects in $N$-body systems undergo irreversible dissipative processes
like those mentioned above, just like less dense matter accumulations,
it is clear that geometric models in general relativity should be
able to account for these types of dynamical behavior of matter sources.
As a first step, it might therefore prove to be a useful idealization
for the treatment of the dissipative phenomena mentioned above that
the associated matter source of the gravitational field does not permanently
remain in a thermal equilibrium state with respect to its environment,
but only at early and late times, so that the existence of HKV, phantom
symmetries and associated metric triples $g_{ab}$, $g'_{ab}$ and
$\tilde{g}_{ab}$ as well as changing degrees of symmetry of spacetime
occur as natural consequences of this particular aspect of Einstein-Hilbert
gravity. After all, given that in everyday life situations it can
be observed that bodies tend to reach a thermal equilibrium state
before and after undergoing dissipative processes, such an idealization
makes far more sense than expecting matter fields to change their
thermal properties throughout the entire temporal evolution of spacetime. 

Yet, as must be clearly emphasized at this point, a complete theoretical
description of the geometrical aspects of dissipative phenomena such
as those mentioned above is extraordinarily difficult, even on a purely
numerical level, and has thereby proved infeasible to date. Consequently,
however, it should be worthwhile to focus instead on a manageable,
yet concrete example of a geometric model with hidden symmetries.
To this effect, one may now proceed by considering the simple case
of Vaidya spacetime, whose metric, in Eddington-Finkelstein coordinates,
can be read off from the line element
\begin{equation}
d\tilde{s}^{2}=-(1-\frac{2m}{r})dv^{2}+2dvdr+r^{2}(d\theta^{2}+\sin^{2}\theta d\phi^{2}).
\end{equation}
In contrast to the Schwarzschild spacetime, the mass function $m(v)$
is time-dependent in the given case, so that Vaidya spacetime is generally
not a vacuum geometry. Rather, it turns out that the energy-moment
tensor of the mentioned geometry characterizes the gravitational field
of an uncharged null-fluid source, which is given by the expression
\begin{equation}
\tilde{G}_{ab}=\frac{2\dot{m}}{r^{2}}dv_{a}dv_{b}.
\end{equation}
Due to the fact that $l_{a}=-dv_{a}$ is an affine geodesic null vector
field, the metric of this spacetime can be cast in Kerr-Schild form,
i.e.
\begin{equation}
\tilde{g}_{ab}=\eta_{ab}+\frac{2m}{r}l_{a}l_{b},
\end{equation}
where $\eta_{ab}$ is the flat Minkowskian metric. By making the specific
choice $m(v):=\chi(v,v_{0})m_{0}$, where $\chi(v,v_{0})$ is a transition
function of the form $(4)$ and $m_{0}=const.$, one then finds that
$m=0$ when $v\leq0$ and $m=m_{0}$ when $v>0$, so that it becomes
clear that line element $(41)$ coincides with that of Minkowski spacetime
for $v\leq0$ and with that of Schwarzschild spacetime for $v>v_{0}$.
The given spacetime therefore represents a toy model which describes
how radiation collapses to form a black hole. 

By adding and subtracting $\frac{2m_{0}}{r}l_{a}l_{b}$ in $(43),$
one therefore obtains two decompositions of $\tilde{g}_{ab}$ of the
form

\begin{equation}
\tilde{g}_{ab}=\eta_{ab}+\frac{2m}{r}l_{a}l_{b}=g_{ab}+\frac{2m'}{r}l_{a}l_{b},
\end{equation}
where $g_{ab}$ is the Schwarzschild metric for mass $m_{0}$ and
$m'=m-m_{0}\leq0$ applies by definition. Thus, in the given case,
one actually obtains a splitting of the form $(30)$ with two spherically
symmetric and static local background metrics $\eta_{ab}$ and $g_{ab}$
that are given with respect to one and the same HKV $\zeta^{a}=\partial_{v}^{a}$,
which is parallel to the Kodama vector field of the geometry. In fact,
due to the fact that the Kodama vector field coincides with $\zeta^{a}$
in the static case (and has all the properties described below), the
Kodama vector field itself can also be identified as HKV.

Anyhow, due to the fact that $L_{\zeta}\tilde{g}_{ab}=2\tilde{\nabla}_{(a}\zeta_{b)}=\frac{2\dot{m}}{r}l_{a}l_{b}$
applies in the given context, one finds that $\tilde{j}^{a}=-\tilde{G}_{\;b}^{a}\zeta^{b}$
is a globally conserved current in the sense that it gives rise to
the conservation law
\begin{equation}
\tilde{\nabla}_{a}\tilde{j}^{a}=0.
\end{equation}
By integrating this relation, one obtains a conserved quasilocal charge
that is zero in $M$ and time independent in $M'$. Using the fact
that $\sqrt{\ensuremath{-\tilde{g}}}=\sqrt{\ensuremath{-g}}=\sqrt{\ensuremath{-\eta}}=r^{2}\sin\theta$
applies for the considered (generalized) Kerr-Schild metrics, from
which it follows that $\sqrt{\tilde{h}}=\frac{r^{2}\sin\theta}{\tilde{N}}=\frac{r^{2}\sin\theta}{\sqrt{1-\frac{2m}{r}}}$
and $d\tilde{\Sigma}^{a}=\frac{r^{2}\sin\theta}{1-\frac{2m}{r}}\partial_{v}^{a}d^{3}x$
applies in the given context, one finds the result

\begin{equation}
\tilde{Q}=8\pi\dot{m}\int\frac{rdr}{r-2m}=8\pi\dot{m}\left[r+2m\ln\vert r-2m\vert\right].
\end{equation}
In the transition region $\mathcal{O}$, this charge coincides with
the Misner-Sharp quasilocal mass and reduces to the ADM mass at spacelike
infinity and the Bondi mass at past and future null infinity (but
at past null infinity it is zero anyway due to the fact that it is
zero in $M$). The fact that the Bondi mass increases with time can
be interpreted as an indication that the gravitating physical system
under consideration absorbs gravitational radiation. Of course, by
choosing instead of $\tilde{j}^{a}=-\tilde{G}_{\;b}^{a}\zeta^{b}$
a current of the form $(38)$, one obtains via $(39)$ also an expression
for the Komar mass, which - due to the validity of conditions $(28)$
and $(29)$ - becomes zero in $M$ and coincides with the Komar mass
of Schwarzschild spacetime in $M'$. The fact that in the present
case a geometric transition between static local spacetimes is considered
has thus, as was to be expected, an influence on the values and forms
of the quasilocal quantities mentioned.

It has to be emphasized that the given example of a geometric transition
between Minkowski space and Schwarzschild spacetime, which describes
the creation of a black hole due to infalling null radiation, is highly
idealized and therefore not of greatest interest from a physical point
of view. However, as will be shown in the following, the developed
methods can also be used to build physically more interesting models
of geometric transitions, such as, for example, one described another
toy model (but a more realistic one) for the geometric transition
between a (two-body) Majumdar-Papapetrou and a Reissner-Nordstr\"om
spacetime, which is to be discussed in the next section. This model
aims to describe the binary merger of two charged black holes, which
remain in a static geometric configuration (described by the two-body
Majumdar-Pappapetrou geometry) for a certain period of time, but then
change their physical properties with time (one of the black holes
accretes null radiation), which forces the entire system to collapse
and thus gives rise to a black hole binary merger. At the end of this
merger process, both black hole singularities coalesce to a single
Reissner-Nordstr\"om black hole singularity, thereby leading to the
formation of a new black hole geometry with associated mass and charge
parameters that are larger than the sum of the masses of the individual
black holes before the beginning of the merger and the associated
geometric transition process. 

What turns out to be special about this particular toy model for a
binary black hole merger is that it shows that, despite the fact that
there are no exact integral laws for the conservation of energy and
angular momentum in black hole mergers (simply because the merger
spacetime has no Killing symmetries at all), a balance equation for
locally conserved quantities can be formulated - in full agreement
with the results presented in this section - which are pretty similar
to the standard conservation laws for the total energy existing in
spacetime with Killing symmetries. As is argued, the balance laws
mentioned therefore represent a reasonable substitute for standard
conservation laws in dynamical spacetimes and, in particular, in black
hole merger geometries.

\section{Black Hole Mergers and Hidden Killing Fields}

After having shown in the previous section that there are spacetimes
in general relativity whose global geometry does not allow for KV,
but does allow for HKV, which have the Killing property only with
respect to the locally symmetric geometry of spacetime, this section
will now discuss a concrete physical example, namely the geometric
model of the binary merger of two extremal Reissner-Nordstr\"om black
holes. 

To describe a binary merger of this kind, a spacetime $(\tilde{M},\tilde{g})$
with manifold structure $\tilde{M}=M\cup\mathcal{O}\cup M'$ shall
be considered, which gives rise to a pair of static spherically local
spacetimes $(M,g)$ and $(M',g')$. The first of these local spacetimes
shall be assumed to describe the static initial configuration of the
geometric field being given by a two-body Majumdar-Papapetrou spacetime
with with two black hole singularities. The components of the corresponding
'initial' metric $g_{ab}$ can be read off the line element

\begin{equation}
ds^{2}=-\frac{dt^{2}}{(1+\frac{m_{1}}{r_{1}}+\frac{m_{2}}{r_{2}})^{2}}+(1+\frac{m_{1}}{r_{1}}+\frac{m_{2}}{r_{2}})^{2}(dx^{2}+dy^{2}+dz^{2}),
\end{equation}
where $m_{i}=e_{i}=const.$ and $r_{i}:=\vert\vec{r}-\vec{r}_{i}\vert=\sqrt{(x-x_{i})^{2}+(y-y_{i})^{2}+(z-z_{i})^{2}}$
with $i=1,2$. By performing a coordinate shift of the type $x\rightarrow x+x_{1}$,
$y\rightarrow y+y_{1}$, $z\rightarrow z+z_{1}$, it can be achieved
that the given line element can be re-written in the form 
\begin{equation}
ds^{2}=-\frac{dt^{2}}{U^{2}}+U^{2}(dr^{2}+r^{2}(d\theta^{2}+\sin^{2}\theta d\phi^{2})),
\end{equation}
using the scalar function $U=1+\frac{m_{1}}{r}+\frac{m_{2}}{\sqrt{r^{2}-2d_{0}\cos\theta r+d_{0}^{2}}}$
which is defined with respect to the relative distance $d_{0}=\sqrt{\vec{d}\vec{d}}=const.$
calculated from the time dependent three-vector $\vec{d}:=\vec{r}_{2}-\vec{r}_{1}$.
As may be checked, this is the same form of the line element as presented
in \cite{chandrasekhar1998mathematical}. 

As may be realized, this initial geometry arises as a specific local
geometric configuration of an associated ambient geometry, namely
that of an ambient spacetime $(\tilde{M},\tilde{g})$ whose metric
components can be read off the line element 

\begin{equation}
d\tilde{s}^{2}=-\frac{dt^{2}}{\tilde{U}^{2}}+\tilde{U}^{2}(dr^{2}+r^{2}(d\theta^{2}+\sin^{2}\theta d\phi^{2})),
\end{equation}
where $\tilde{U}(t,r,\theta)=1+\frac{m_{1}+m_{0}}{r}+\frac{m_{2}}{\sqrt{r^{2}-2d\cos\theta r+d^{2}}}$
applies by definition. While the mass scalars $m_{1}$ and $m_{2}$
occurring in $\tilde{U}(t,r,\theta)$ are the same as in $(48)$,
the remaining scalar functions occurring in the metric components
of the ambient spacetime $(\tilde{M},\tilde{g})$ have the form $m_{0}(t,t_{1})=\chi(t,t_{1})m_{0}$
with $m_{0}=const.$ and $d(t,t_{2})=\left(1-\chi(t,t_{2})\right)d_{0}$. 

After having now obtained this form of the line element of the ambient
spacetime $(\tilde{M},\tilde{g})$, it is necessary to show that the
considered toy model is consistent with the geometric deformation
approach treated in the previous section. This can easily be done.
All one has to do is to add and subtract (in a fixed coordinate chart)
the components of the 'initial' Majumdar-Papapetrou metric and the
'final' extremal Reissner-Nordstr\"om metric to/from that of the corresponding
ambient metric. By taking these steps, it then becomes clear that
the corresponding metric can be written in the form
\begin{equation}
\tilde{g}_{ab}=g{}_{ab}+\mathrm{e}_{ab}=g'_{ab}+\mathrm{e}'_{ab},
\end{equation}
where $g_{ab}$ is the Majumdar-Papapetrou metric and $g'_{ab}$ is
the extremal Reissner-Nordstr\"om metric of the black hole resulting
from the merger process. Since for the given choice for $\mathrm{e}_{ab}$
and $\mathrm{e}'_{ab}$ it becomes clear that $\text{\ensuremath{\mathrm{e}_{ab}}}\vert_{M}=0$
and $\mathrm{e}'_{ab}\vert_{M'}=0$ such that $g_{ab}\equiv\tilde{g}_{ab}\vert_{M}$
and $g'_{ab}\equiv\tilde{g}_{ab}\vert_{M'}$, all circumstances are
the same as required in the first chapter. 

The idea behind considering this type of ambient spacetime is the
following: Due to the properties of $\chi(t,t_{i})$ with $i=1,2$,
it becomes clear that the expression $m_{0}(t,t_{1})$ is zero for
$t\leq0$. After this period, said expression starts to grow until
it reaches a maximum value of $m_{0}$ in $t>0$. Consequently, however,
it becomes clear that the mass of the first black hole increases in
said time period from a value of $m_{1}=const.$ to the larger value
$m_{1}+m_{0}=const.$, so that it can be concluded that one of the
two black holes continuously accretes matter and thereby becomes more
massive. The corresponding accretion process, which (by assumption)
takes place in the time interval $[0,t_{1}]$, perturbs and disrupts
the stable static equilibrium of the two charged black holes. As a
result of this disruption, it can be expected that the black hole
binary system slowly becomes unstable from a fixed point of time onwards.
In the given toy model, which unfortunately is not able to take into
account the inspiral which binary black hole systems are expected
to undergo under realistic circumstances, this has the sole consequence
that the distance between the black hole singularities becomes smaller
and smaller due geometric fluctuations that are caused by an increase
of the mass of one of the black holes. As may be expected, this process
continues until both black hole singularities 'collide' (in the sense
that they reach the same position) and 'coalesce' to a single black
hole (in the sense that the binary black hole geometry becomes that
of a single Reissner-Nordstr\"om black hole). Consequently, however,
the binary merger process must lead to a static geometric configuration,
that is, a local spacetime $(M',g')$, whose metric components can
be read off the Reissner-Nordstr\"om line element

\begin{equation}
ds'^{2}=-(1-\frac{2m'}{r}+\frac{e'^{2}}{r})dt^{2}+\frac{dr^{2}}{1-\frac{2m'}{r}+\frac{e'^{2}}{r}}+r^{2}(d\theta^{2}+\sin^{2}\theta d\phi^{2})),
\end{equation}
where $m'=m_{1}+m_{2}+m_{0}>m_{1}+m_{2}$ applies in the present context.
As can readily be checked, line element $(48)$ reduces to the form
\begin{equation}
ds'^{2}=-\frac{dt^{2}}{U'^{2}}+U'^{2}(dr^{2}+r^{2}(d\theta^{2}+\sin^{2}\theta d\phi^{2}))
\end{equation}
in $M'$, where $U'=1+\frac{m'}{r}$ applies by definition. Thus,
by using now the fact that the black hole geometric configuration
resulting from the merger is still an extremal Reissner-Nordstr\"om
black hole for which $m'=e'$ applies, it is not difficult to see
that the introduction of a new radial coordinate $r'=r+m'$ allows
one to re-write the corresponding line element in the form
\begin{equation}
ds^{2}=-(1-\frac{m'}{r'})^{2}dt^{2}+\frac{dr'^{2}}{(1-\frac{m'}{r'})^{2}}+r'^{2}(d\theta^{2}+\sin^{2}\theta d\phi^{2})),
\end{equation}
which, however, is obviously identical to line element $(51)$. 

With that settled, it may be noted that the Einstein tensor of this
ambient spacetime decomposes according to the rule
\begin{equation}
\tilde{G}_{ab}=G_{ab}+\rho_{ab}=G'_{ab}+\rho'_{ab},
\end{equation}
where $G_{ab}$ is the Majumdar-Papapetrou Einstein tensor, $G'_{ab}$
is the Reissner-Nordstr\"om Einstein tensor and $\rho_{ab}$ and $\rho'_{ab}$
are the dynamical parts of Einstein tensor $\tilde{G}_{ab}$ of the
merger geometry. The first term $G_{ab}+\rho_{ab}$ has the simple
property that it must necessarily agree with the Einstein-Tensor of
the Majumdar-Papapetrou geometry in $M$, i.e.
\begin{equation}
\tilde{G}_{ab}\vert_{M}=G_{ab}=8\pi T_{ab},
\end{equation}
where the corresponding energy-momentum tensor consists of a dust
part and a part characterizing the electromagnetic field, i.e.
\begin{equation}
T_{ab}=\varepsilon u_{a}u_{b}+F_{ac}F_{\;b}^{c}-\frac{1}{4}g_{ab}F_{cd}F^{cd}.
\end{equation}
Here, $\varepsilon$ represents the energy density and $u^{a}$ the
four velocity of the two-body system and $F_{ab}=2\nabla_{[a}A_{b]}$
is the Maxwell tensor being defined with respect to the vector potential
$A_{b}=\frac{1}{U}\partial_{t}^{a}$. To reobtain the associated Einstein
tensor of the corresponding local Majumdar-Papapetrou spacetime $(M,g)$
from that of the ambient spacetime $(\tilde{M},\tilde{g})$ given
by expression $(54)$, is becomes clear that the dynamical part of
the total Einstein tensor $\tilde{G}_{ab}$ of the merger geometry
must be zero in $M$, i.e. $\rho_{ab}\vert_{M}=0.$ Of, course, the
same situation occurs also late times with respect to the local Reissner-Nordstr\"om
spacetime $(M',g')$, that is in $M'$, where there must hold

\begin{equation}
\tilde{G}_{ab}\vert_{M'}=G'_{ab}=8\pi T'_{ab}
\end{equation}
and thus $\rho_{ab}\vert_{M'}=0$ as a direct consequence. The corresponding
stress-energy tensor of the merger spacetime thus reads 
\begin{equation}
T'_{ab}=F'_{ac}F_{\;b}^{'c}-\frac{1}{4}g'_{ab}F'_{cd}F^{'cd}.
\end{equation}
at late times, from which it can be concluded that both the initial
and final geometric configurations prove to be static, spherical spacetimes,
whereas the complete merger geometry of spacetime has none of these
geometric properties; rather, it proves to be obviously non-stationary
and axisymmetric. Here, it may be noted that much about the geometrical
structure of axisymmetric solutions of Einstein's field equations
is known in the literature. Accordingly, based on the results of \cite{chandrasekhar1998mathematical}
(see pages $142/143$), one finds after appropriate identification
of the occurring terms that the only non-vanishing components of $\rho_{ab}$
are
\begin{align}
\rho_{tt}= & \:3(\partial_{t}\tilde{U})^{2}-3(\partial_{t}U)^{2}+\frac{\tilde{U}^{4}\left(2r^{2}U\partial_{r}^{2}U-3r^{2}(\partial_{r}U)^{2}+U^{2}\right)}{r^{2}\tilde{U}^{4}U^{4}}-\\
- & \frac{U^{4}\left(2r^{2}\tilde{U}\partial_{r}^{2}\tilde{U}-3r^{2}(\partial_{r}\tilde{U})^{2}+\tilde{U}^{2}\right)-\tilde{U}^{4}\left(2U\partial_{\theta}^{2}U-(\partial_{\theta}U)^{2}+2U\partial_{\theta}U\cot\theta\right)}{r^{2}\tilde{U}^{4}U^{4}}-\nonumber \\
- & \frac{U^{4}\left(2\tilde{U}\partial_{\theta}^{2}\tilde{U}-(\partial_{\theta}\tilde{U})^{2}+2\tilde{U}\partial_{\theta}\tilde{U}\cot\theta\right)}{r^{2}\tilde{U}^{4}U^{4}},\nonumber \\
\rho_{rr}=\: & 2\partial_{t}^{2}UU+3(\partial_{t}U)^{2}-2\partial_{t}^{2}\tilde{U}\tilde{U}-3(\partial_{t}\tilde{U})^{2}-\\
- & \frac{\tilde{U}^{4}\left(r^{2}(\partial_{r}U)^{2}+(\partial_{\theta}U)^{2}\right)-U^{4}\left(r^{2}(\partial_{r}\tilde{U})^{2}+(\partial_{\theta}\tilde{U})^{2}\right)}{r^{2}\tilde{U}^{4}U^{4}},\nonumber \\
\rho_{\theta\theta}= & \rho_{\phi\phi}=2\partial_{t}^{2}\tilde{U}\tilde{U}+(\partial_{t}\tilde{U})^{2}-2\partial_{t}^{2}UU-(\partial_{t}U)^{2}+\\
+ & \frac{U^{4}\left(r^{2}(\partial_{r}\tilde{U})^{2}-(\partial_{\theta}\tilde{U})^{2}\right)+\tilde{U}^{4}\left(r^{2}(\partial_{r}U)^{2}-(\partial_{\theta}U)^{2}\right)}{r^{2}\tilde{U}^{4}U^{4}},\nonumber 
\end{align}
where the results 
\begin{align}
\partial_{t}\tilde{U}= & \frac{\dot{m}_{0}}{r}-\frac{m_{2}\dot{d}(d-r\cos\theta)}{\left(r^{2}-2rd\cos\theta+d^{2}\right)^{\frac{3}{2}}},\\
\partial_{t}^{2}\tilde{U}= & \frac{\ddot{m}_{0}}{r}+\frac{3m_{2}\dot{d}^{2}(d-r\cos\theta)^{2}}{\left(r^{2}-2rd\cos\theta+d^{2}\right)^{\frac{5}{2}}}-\frac{m_{2}\left(\ddot{d}(d-r\cos\theta)+\dot{d}^{2}\right)}{\left(r^{2}-2rd\cos\theta+d^{2}\right)^{\frac{3}{2}}},\\
\partial_{r}\tilde{U}= & -\frac{m_{0}+m_{1}}{r^{2}}-\frac{m_{2}(r-d\cos\theta)}{\left(r^{2}-2rd\cos\theta+d^{2}\right)^{\frac{3}{2}}},\\
\partial_{r}^{2}\tilde{U}= & \frac{2(m_{0}+m_{1})}{r^{3}}+\frac{3m_{2}(r-d\cos\theta)^{2}}{\left(r^{2}-2rd\cos\theta+d^{2}\right)^{\frac{5}{2}}}-\frac{m_{2}}{\left(r^{2}-2rd\cos\theta+d^{2}\right)^{\frac{3}{2}}},\\
\partial_{\theta}\tilde{U}= & -\frac{m_{2}dr\sin\theta}{\left(r^{2}-2rd\cos\theta+d^{2}\right)^{\frac{3}{2}}},\\
\partial_{\theta}^{2}\tilde{U}= & \frac{3m_{2}d^{2}r^{2}\sin^{2}\theta}{\left(r^{2}-2rd\cos\theta+d^{2}\right)^{\frac{5}{2}}},
\end{align}
as well as the relations $\partial_{t}U=\partial_{t}^{2}U=0$ have
been used. Since one also has $\partial_{t}U'=\partial_{t}^{2}U'=0$,
the non-vanishing components of $\rho'_{ab}$ can simply be obtained
by replacing $U$ by $U'$ in $(59-67)$. Using then the fact that
$\dot{\chi}\propto\chi$ and $\ddot{\chi}\propto\chi$ one finds that
$\partial_{t}\tilde{U}\vert_{M}=\partial_{t}U$, $\partial_{t}^{2}\tilde{U}\vert_{M}=\partial_{t}^{2}U$,
$\partial_{t}\tilde{U}\vert_{M'}=\partial_{t}U'$, $\partial_{r}^{2}\tilde{U}\vert_{M'}=\partial_{r}^{2}U'$,
$\partial_{r}\tilde{U}\vert_{M}=\partial_{r}U$, $\partial_{r}^{2}\tilde{U}\vert_{M}=\partial_{r}^{2}U$,
$\partial_{r}\tilde{U}\vert_{M'}=\partial_{r}U'$, $\partial_{r}^{2}\tilde{U}\vert_{M'}=\partial_{r}^{2}U'$,
and $\partial_{\theta}\tilde{U}\vert_{M}=\partial_{\theta}U$, $\partial_{\theta}^{2}\tilde{U}\vert_{M}=\partial_{\theta}^{2}U$,
$\partial_{\theta}\tilde{U}\vert_{M'}=\partial_{\theta}U'=0$, $\partial_{\theta}^{2}\tilde{U}\vert_{M'}=\partial_{\theta}^{2}U'=0$,
which implies that $\rho_{ab}\vert_{M}=0$ and $\rho_{ab}\vert_{M'}=0$
applies as desired.

Having clarified that, it may next be noted that the Majumdar-Papapetrou
metric $g_{ab}$ given by line element $(48)$ coincides locally with
that of an one-body extremal Reissner-Nordstr\"om black hole system
in the limit $\text{\ensuremath{d_{0}\rightarrow\infty}}$. A multipole
expansion can therefore be used in $M$ to describe perturbations
of this metric due to the second black hole for large enough $d_{0}$,
i.e. in the case of large spatial separation. Since one of the black
holes starts to constantly accrete matter in $\tilde{M}\backslash M$,
these perturbations grow larger and larger over time, so that the
system slowly becomes unstable in the sense that the distance between
the singularities of the black holes becomes smaller and smaller as
time goes by. This destabilization process continues until both black
holes collide and merge with each other. Consequently, however, in
order to indicate that a two-body system evolves into a one-body system
in the process, it makes sense to re-write metric decomposition relation
$(50)$ given above in the form 
\begin{equation}
\tilde{g}_{ab}=g^{(j)}{}_{ab}+\mathrm{e}_{ab}^{(j)}=g{}_{ab}^{(j-1)}+\mathrm{e}{}_{ab}^{(j-1)},
\end{equation}
where the index $j$ counts the number of bodies in the system, which,
in the given case, is just two. Moreover, in this very relation, which
is completely identical to decomposition relation $(50),$ the tensor
$g^{(j)}{}_{ab}$ denotes the local Majumdar-Papapetrou metric, which
reduces in the limit of infinite spatial separations, i.e. in the
limit $\text{\ensuremath{d_{0}\rightarrow\infty}}$ to the local metric
of an asymptotic extremal Reissner-Nordstr\"om black hole either with
mass and charge $m_{1}=e_{1}$ or with mass and charge $m_{2}=e_{2}$.
On the other hand, $g_{ab}^{(j-1)}$ is the metric of the extremal
Reissner-Nordstr\"om black hole mass and charge $m'=e'$, which occurs
only after the black hole merger has taken place. This local form
of the ambient metric $\tilde{g}_{ab}$ is relevant not least because
the deformation field $\mathrm{e}{}_{ab}^{(j-1)}=g_{ab}-g^{(j-1)}{}_{ab}$
is zero in the local subregion $M'\subset\tilde{M}$. In turn, the
deformation tensor fields $\mathrm{e}_{ab}^{(j)}=g_{ab}-g^{(j)}{}_{ab}$
become zero in the local subregion $M\subset\tilde{M}$, where the
ambient spacetime $(\tilde{M},\tilde{g})$ coincides with the local
Majumdar-Papapetrou configuration $(M,g)$. In the complement of this
region, i.e. in $\tilde{M}\backslash M$, the merger geometry $(\tilde{M},\tilde{g})$
describes how perturbations caused by material accreted by one of
the black holes lead to a binary merger of the black holes, whereas,
at the end of this merging process, a single Reissner-Nordstr\"om black
hole is formed whose mass is larger than the sum of the masses of
the individual black holes before the beginning of the geometric transition
process. That is to say, the merger spacetime $(\tilde{M},\tilde{g})$
reaches again a static spherically symmetric geometric configuration
at late times, i.e. in $M'\subset\tilde{M}$, where the geometry of
the ambient spaectime $(\tilde{M},\tilde{g})$ coincides with that
of $(M',g')$. 

Consequently, however, although the geometric model considered thus
far is only a Mickey Mouse toy model, it turns out to be fully consistent
with Hawking's area theorem, which states that when two black holes
merge, the area of the final event horizon is greater than or equal
to the sum of the areas of the initial horizons. Moreover, since the
two-body system settles down to a stationary spherically symmetric
state after the black hole merger, the considered toy model is also
consistent with the black hole uniqueness theorems and the no hair
theorems \cite{Carter:1971zc,hawking1972black,Israel:1967wq,Israel:1967za,Mazur:1982db,Robinson:1974nf,Robinson:1975bv}. 

To see that not only the latter, but also the first part of this statement
is true, it may be checked that the energy-momentum tensor of the
merger geometry meets the weak null energy condition 
\begin{equation}
\tilde{T}_{ab}K^{a}K^{b}\geq0
\end{equation}
for every null vector $K^{a}$. In addition, one may consider line
element $(47)$, introduce a radial coordinate $\bar{r}=r+m_{1}$
and calculate the expression 
\begin{equation}
\mathcal{A}_{1}=\underset{\bar{r}\rightarrow m_{1}}{\lim}\overset{2\pi}{\underset{0}{\int}}\overset{\pi}{\underset{0}{\int}}\sqrt{g_{\theta\theta}g_{\phi\phi}}d\theta d\phi=4\pi m_{1}^{2}.
\end{equation}
Then, by taking into account that all foregoing remarks made with
respect to $m_{1}$ can also be made with respect to $m_{2}$ (which
is not least because each point $r_{i}$ with $i=1,2$ in line element
$(47)$ represents a null hypersurface with an area $4\pi m_{i}$
\cite{chandrasekhar1998mathematical}), one may transform line element
$(48)$ in suitable coordinates and calculate the further expression
\begin{equation}
\mathcal{A}_{2}=\underset{\bar{\bar{r}}\rightarrow m_{2}}{\lim}\overset{2\pi}{\underset{0}{\int}}\overset{\pi}{\underset{0}{\int}}\sqrt{g_{\theta\theta}g_{\phi\phi}}d\theta d\phi=4\pi m_{2}^{2}.
\end{equation}
Sure enough, the expressions obtained can be combined, i.e.
\begin{equation}
\mathcal{A}_{1}+\mathcal{A}_{2}=4\pi(m_{1}^{2}+m_{2}^{2}).
\end{equation}
Using next line element $(53)$, which coincides with line element
$(51),$ one obtains the result
\[
\mathcal{A}'=\underset{r'\rightarrow m'}{\lim}\overset{2\pi}{\underset{0}{\int}}\overset{\pi}{\underset{0}{\int}}\sqrt{g'_{\theta\theta}g'_{\phi\phi}}d\theta d\phi=4\pi m'^{2}.
\]
However, since it is known that $m'>m_{1}+m_{2}$, this allows one
to conclude that Hawking's area theorem
\begin{equation}
\mathcal{A}_{1}+\mathcal{A}_{2}<\mathcal{A}'
\end{equation}
holds true in the given case even when $m_{0}\ll m_{1},m_{2}$ and
therefore $m'^{2}\simeq(m_{1}+m_{2})^{2}$ applies, which happens
to be a case of greatest interest from a physical point of view. It
may be noted that exactly the same result was obtained previously
in \cite{astorino2020enhanced}.

To proceed, one may take the fact into account that the vector field
$\zeta^{a}=\partial_{t}^{a}$ living on the ambient spacetime $(\tilde{M},\tilde{g})$
coincides with the KV of the local spacetimes $(M,g)$ and $(M',g')$.
Clearly, this vector field also defines a Papapetrou field in the
sense of \cite{bini2004simon,fayos1999papapetrou}. Consequently,
however, it can be concluded that $\zeta^{a}$ is also a HKV in the
sense of the previous section satisfying condition $(32)$ and that
the ambient spacetime $(\tilde{M},\tilde{g})$ has phantom symmetry.
One is therefore in the favorable situation that the corresponding
current $\tilde{j}^{a}=\tilde{G}_{\;b}^{a}\zeta^{b}$ can be used
to formulate an integral law, which takes the form $(33)$ in $M$
and $(34)$ in $M'$. To see this, one may consider in the same way
as in the first section two submanifolds $\mathcal{D}\subset M$ and
$\mathcal{D}'\subset M'$ with boundaries $\partial\mathcal{D}$ and
$\partial\mathcal{D}'$, which are contained in an associated ambient
submanifold $\tilde{\mathcal{D}}\subset\tilde{M}$ with boundary $\partial\mathcal{\tilde{D}}$
such that $\mathcal{D},\mathcal{D}'\subseteq\tilde{\mathcal{D}}$.
This geometric setting can be used to deduce the relation 
\begin{equation}
\underset{\mathcal{\tilde{D}}}{\int}\tilde{\nabla}_{a}\tilde{j}^{a}d\tilde{v}=2\underset{\mathcal{\tilde{D}}}{\int}\tilde{G}_{ab}\tilde{\nabla}^{(a}\zeta^{b)}d\tilde{v}.
\end{equation}
Using Gauss' theoreom and decomposition relation $(35)$ to convert
the LHS of $(73)$, one thus obtains a balance law of the form $(37)$,
in relation to which $L_{\zeta}Q=L_{\zeta}Q'=0$ applies. However,
this implies that also $L_{\zeta}\mathscr{G}=0$ applies, from which
it can be concluded that a local observer witnessing a geometric transition
from $(M,g)$ to $(M',g')$ with respect to the ambient spacetime
$(\tilde{M},\tilde{g})$ could come to the conclusion that the total
energy of the system is conserved, although the ambient spacetime
lacks the corresponding Killing symmetry! But this seems to suggest
that the derived balance law could retain its validity not only in
the considered, extremely simplified Mickey Mouse model of a binary
black hole merger, but also in more realistic modelings of merger
geometries. For this reason, HKV appear to constitute good and reasonable
substitutes for KV dynamical spacetimes and, in particular, in black
hole merger geometries.

\section*{Conclusion}

In the present work, the geometrical properties of special classes
of spacetimes have been studied, whose geometry is not globally, but
only locally symmetric, and which therefore admit Killing symmetries
only in a local subregion of spacetime, not in the entire Lorentzian
manifold. Such spacetimes, as it has been argued, have the advantageous
property that they allow one to specify a preferred class of vector
fields, the class of so-called HKV, which are geometrically distinguished
by coinciding exactly with associated Killing vector fields at local
scales, i.e., in the local region where spacetime becomes symmetric.
This particular property has been exploited in the present work to
show that certain integral laws of the theory reduce to exact conservation
laws on local scales, thereby leading to exact expressions for energy
and angular momentum in the symmetric parts of spacetime. Using this
insight, physical balance laws for energy and angular momentum were
derived, which describe how exactly any of these locally conserved
quantities change in time during the evolution of a generic ambient
geometry that completely lacks any sort of Killing symmetry. The derived
balance laws have the interesting property that they are very closely
related to the integral conservation laws of known symmetric models
and, except for the correction terms resulting from the geometric
deformation of the symmetric local background geometry, coincide with
these conservation laws even in the face of a generic geometric setting.
In order to demonstrate the utility of the laws mentioned, a special
geometric model was treated in the second part of this work, namely
a toy model for a merger of two extremal Reissner-Nordstr\"om black
holes. For the exact characterization of the geometric structure of
a merger geometry of this type, a non-stationary axially symmetrical
geometric model was considered, which, by construction, coincides
locally with the (two-body) Majumdar-Papapetrou solution at early
times and the Reissner-Nordstr\"om solution at later times. For this
model, the existence of HKV has been demonstrated, and it has been
proved that the geometric structure of the merger spacetime proves
to be consistent with Hawking's area theorem and also allows the definition
of locally conserved integral expressions, which could prove to be
useful for the generalization of the laws of black hole mechanics
for merger geometries in the future.
\begin{description}
\item [{Acknowledgements:}]~
\end{description}
I want to thank Felix Wilkens for his support in preparing the images
depicted in Figure 1 of the paper.

\bibliographystyle{plain}
\addcontentsline{toc}{section}{\refname}\nocite{*}
\bibliography{1C__Arbeiten_Papers_litphant}

\end{document}